# Enhancing Navigation on Wikipedia with Social Tags

Arkaitz Zubiaga

*NLP&IR Group, UNED, Madrid, Spain*

**SUMMARY.**

**Social tagging has become an interesting approach to improve search and navigation over the actual Web, since it aggregates the tags added by different users to the same resource in a collaborative way. This way, it results in a list of weighted tags describing its resource. Combined to a classical taxonomic classification system such as that by Wikipedia, social tags can enhance document navigation and search. On the one hand, social tags suggest alternative navigation ways, including pivot-browsing, popularity-driven navigation, and filtering. On the other hand, it provides new metadata, sometimes uncovered by documents' content, that can substantially improve document search. In this work, the inclusion of an interface to add user-defined tags describing Wikipedia articles is proposed, as a way to improve article navigation and retrieval. As a result, a prototype on applying tags over Wikipedia is proposed in order to evaluate its effectiveness.**

## 1. INTRODUCTION

Social tagging has become an interesting approach to improve search and navigation over the actual Web. The success of social bookmarking sites such as Delicious[1] and StumbleUpon[2] has shown the usefulness of social tags as document metadata to ease the subsequent information retrieval and navigation.

A social tagging system aggregates the tags assigned by different users to the same resource in a collaborative way. After a set of users annotate a resource with their preferred tags, it results in a list of weighted tags describing the resource. It has been demonstrated that all these tags, used as metadata, improve information retrieval (Heymann et al. 2008) and navigation (Millen and Feinberg 2006). The main reason for its success is the open vocabulary it is based on, since it makes possible the use of previously unused tags. This causes a wide variety of metadata to be available for each resource, and so it extends the possible ways to retrieve a resource. In a set of experiments over a collection of user-annotated web documents, Noll and Meinell (2008) show a high number of new tags, not appearing in document's content, to be available for popular documents. Another reason for the success of social tagging systems is the ability to aggregate tags added by different users, providing a list of weighted tags.

---

[1] http://delicious.com
[2] http://www.stumbleupon.com

Article navigation and search on Wikipedia is one of the most important issues to successfully access to the information it offers. The free encyclopedia provides the following navigation methods:

- **Search engine:** the search engine provided by MediaWiki adds the possibility to search articles by means of keywords, like in any search engine, such as Google. This is a good approach to look for an unknown article name. Anyway, the lack of many words in articles' content may difficult the retrieval.
- **Category-driven navigation:** the taxonomy of Wikipedia allows to browse articles sharing the same category, and going through parent and child categories. Though this is very interesting to organize all the content in Wikipedia, a taxonomy is always limited to the defined categories and, on the other hand, an article may only be present or not in a category, without different weights on that relation.
- **Link-driven navigation:** one of the main characteristics of Wikipedia is the high number of links its articles usually have. These links provide the way to navigate through related articles. Anyway, navigating this way is subject to link availability, and the fact of not finding the desired article could make us think it does not exist.

So there are multiple ways to navigate through Wikipedia, but could social tagging improve article search and retrieval? This issue is covered in this work.

## 2. SOCIAL TAGGING

Social tagging has become an effective approach to annotate documents and other items since its early implementations on sites such as the well-known social bookmarking site Delicious. Tags are an open way to assign keywords to an item, in order to describe it and to ease its subsequent retrieval. As opposed to a classical taxonomic categorization system, they are usually non-hierarchical, and the vocabulary is fully open, so users are free to define the desired keywords. For instance, a user could tag Wikipedia as *free*, *encyclopedia* and *web2.0* whereas another user could use *collaborative* and *information* tags to annotate it.

A tagging system becomes social when these tags are publicly visible, and so profitable for anyone. A user could take advantage of tags defined by others to retrieve a resource, e.g., a web site.

Tagging systems can, generally, be categorized into two types:

- **Simple tagging:** Only the owners of the resource add tags to it. For instance, Flickr[3] can be considered a simple tagging system, where only the user uploading an image/photo tags it.

---

[3] http://www.flickr.com

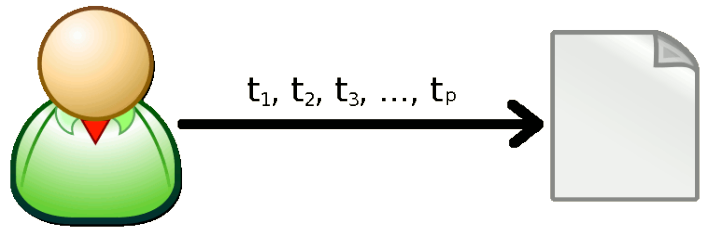

*Figure 1: Example of a simple tagging system*

- **Collaborative tagging:** All the users can tag a resource, not only its author. Generally, tags are defined by resources' users, and as a result of many users tagging the same item, a weighted set of tags is available for each resource. For instance, Delicious, as a social bookmarking site, is a collaborative tagging system, where each resource (URL) could be annotated (tagged) by as many users as considered it interesting.

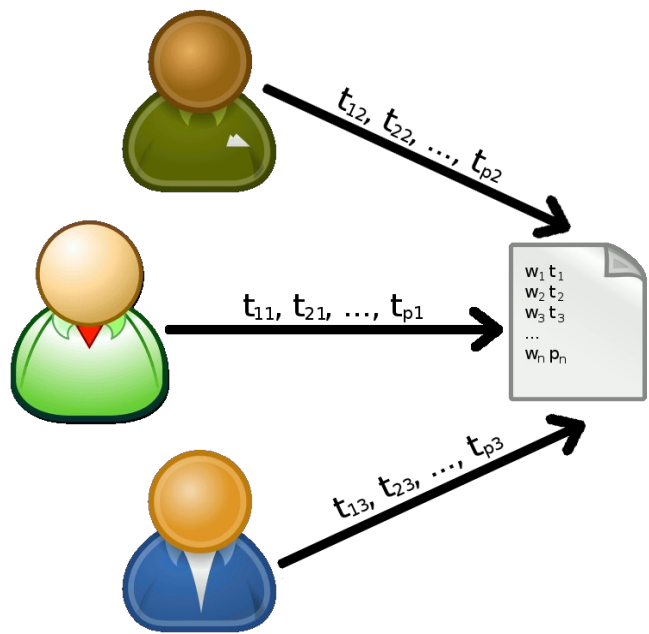

*Figure 2: Example of a collaborative tagging system*

As a result of many users tagging resources, social tagging sites present a tag cloud. A tag cloud is simply a list of the most popular tags in the site. Tags in this list usually have different font sizes, where the bigger font size the more resources it has in it. The Figure 3 shows a tag cloud on Delicious, where noticeably *blog* and *design* tags are the most popular in the site.

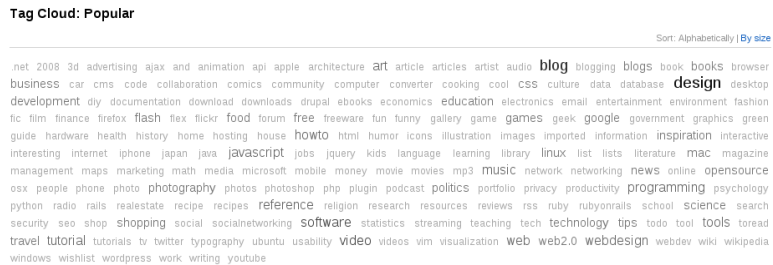

*Figure 3: Example of a tag cloud (Extracted from Delicious)*

## 3. INCLUDING TAGS INTO NAVIGATION

The navigation methods offered by Wikipedia shown above are very interesting, and complementary as well. Nonetheless, it may sometimes be insufficient to find the desired article. Classical information repositories such as the Open Directory Project[4] (ODP) rely on a top-down categorization system, whereas other sites like Amazon[5] successfully combine it with a bottom-up social tagging system. Amazon believes that social tags can substantially fulfill search keywords by users, sometimes uncovered by the taxonomy or the resource's content itself. Besides the classical navigation by means of articles pertaining to a category in a taxonomic system, social tags suggest alternative ways (Smith 2007):

- **Pivot-browsing:** this means moving through an information space by choosing a reference point to browse, e.g., pivoting on a tag allows to look for related tags.
- **Popularity-driven navigation:** sometimes a user would like to get documents that are popular for a known tag, e.g., retrieving only the documents where a tag has been top-weighted.
- **Filtering:** social tagging allows to separate the stuff you do not want from the stuff you do want, e.g., gathering documents containing a tag but excluding another one.

### 3.1 Benefits of Tagging

Social tags may enhance the navigation and search through huge collections of information like Wikipedia. Some of the advantages of social tags are the following:
- It is very simple for a user to define tags for an article. A user can define some keywords, as if they were the keywords he would use to look for the article in a search engine.

---

[4] http://www.dmoz.org
[5] http://www.amazon.com

- Tags rely in an open vocabulary, and so it may continue increasing in order to adjust to the needs of new data. A user does not need to check which tags already exist, unlike with categories.
- Each user can assign different tags to the same resource. This allows to create a list of weighted tags for each article, aggregating all users' tags. This means that each tag is not only present or not for each article, but it may have a weight of relevance, indicating how relevant the tag is for the article.

### 3.2 Enhancing Taxonomies by means of Tagging

The features of both categories and tags have been shown above. However, it is really important to evaluate whether a system with categories could be improved by using tags. The main advantages of adding tags to a taxonomic system are:
- Providing the aforementioned new ways to navigate.
- Providing new terms to describe the articles, and so the search engine could rely in a higher amount of words to search on.
- Providing a bottom-up classification, instead of a top-down one.

### 3.3 Avoiding Tag Mess

On the other hand, a tagging system also has its drawbacks. The fact of relying on an open vocabulary could also be a disadvantage if it is not handled as it should. A user could define a tag like "science fiction", while other could tag as "science-fiction" or "sci-fi". If the system is able to merge all these tags, and consider them as if they were the same, the results would be better. To achieve this, a great solution is to rely on a method like that by Librarything[6]. This site allows users to define relations between tags, indicating that some of them have the same meaning. A huge community, like that by Wikipedia, could work on relating tags and improving the organization to avoid tag mess.

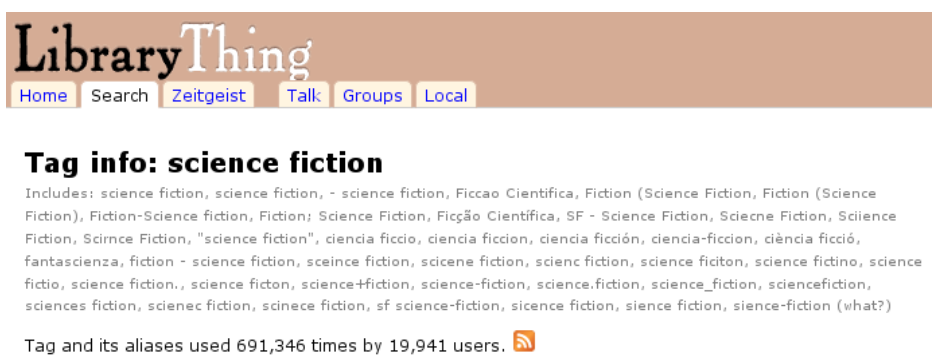

*Figure 4: An example of related tags on Librarything*

---

[6] http://www.librarything.com

## 4. EXPERIMENTS

In order to carry out a set of experiments to evaluate whether social tagging could help improving navigation and search through Wikipedia, a prototype on applying tags over the English version of the free encyclopedia has been developed. This prototype uses tag information collected from Delicious, as explained below. The prototype is available on the web[7].

### 4.1 Dataset Generation

To create the prototype on tags over Wikipedia, a set of tags for the English version of the free encyclopedia was gathered. Starting with a set of more than 2 million articles from the English Wikipedia on April 2009, the tag information for each of these articles was retrieved from the social bookmarking site Delicious. All this tag information included some non-relevant tags within the Wikipedia environment, which were removed from the list. These tags were *wikipedia*, *reference* and *wiki*. These tags seem to be interesting to describe the Wikipedia itself, but not each of its articles. Finally, only the articles annotated by at least 10 users in Delicious were preserved.

As a result, a dataset with 20,764 tagged Wikipedia articles was generated. This dataset is available for download[8].

### 4.2 Results

The results of using this tag data over Wikipedia can be evaluated in two ways. On the one hand, in a qualitative way, by means of navigating through the prototype mentioned above. This prototype allows us to evaluate whether it is helpful to improve article navigation and retrieval. The new navigation ways by means of pivot-browsing, filtering and popularity-based navigation seem to be really interesting.

On the other hand, a quantitative analysis can also be done by means of analyzing tag data. Aggregating all the tags with their weights from all the articles generated a tag cloud where some tags like *programming*, *psychology*, *science*, *research*, *software* and *architecture* stand out (see Figure 5).

---

[7] http://taggedwiki.zubiaga.org
[8] http://nlp.uned.es/social-tagging/

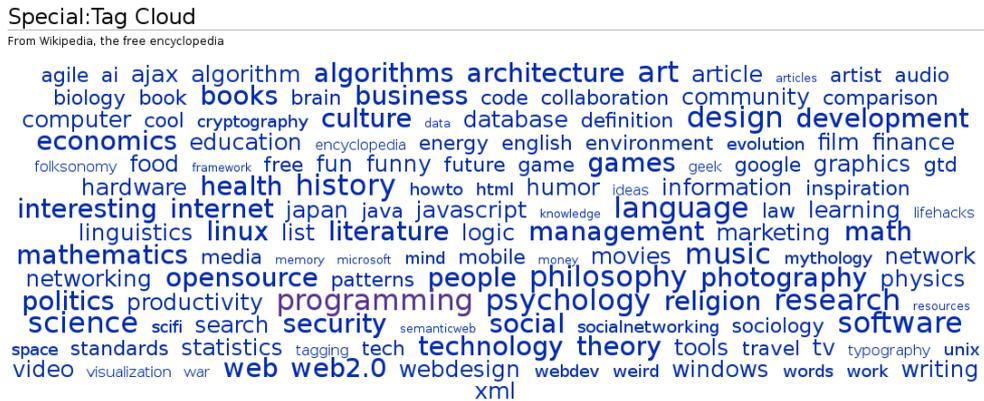

*Figure 5: The resulting tag cloud for the experiments*

Taking into account the occurrences of the tags within the articles, different views could be applied. The occurrence of a tag can be checked within an article's content, within the category names where the article is classified, or even within the whole document (both categories and content) (see Table 1). It is noteworthy that tags are only present in less than 50% of the articles' content, and less than 8% of the categories. This shows how lots of new terms occurred for each article, simply by using social tags.

On the other hand, the occurrence of the tags within the articles varies depending on the number of tags the article has. An article from Wikipedia annotated on Delicious could only have 2, 3 or 4 different tags, while another one could have 30 different tags. An analysis separating the articles depending on the number of tags they have, shows how the more tags an article has, the more new terms (tags) emerge (see Figure 6).

|  | **Found** | **Not Found** | **% Found** |
|---|---|---|---|
| **Document** | 251,139 | 206,569 | 54.86% |
| **Content** | 202,151 | 255,557 | 44.16% |
| **Categories** | 35,237 | 422,471 | 7.70% |

*Table 1: Tag presence stats for the experiments*

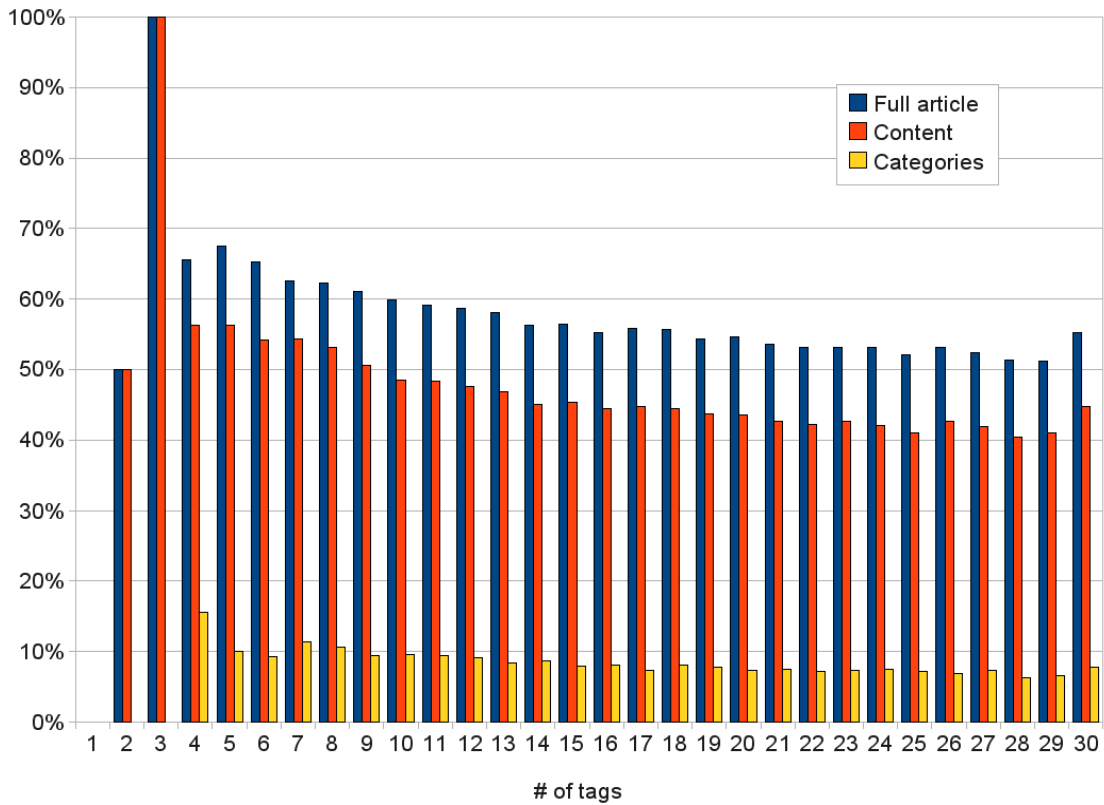

*Figure 6: Percent of tags presence, depending on the number of tags its article has*

Finally, some examples show that there are lots of articles not containing an apparently related term on its content, but contained within its set of tags. Some examples are:
- Users defined the tag *programming* for the article *Framework*, but that word is not present in the document.
- The same occurs for the tag *mathematics* in the article *Zipf's law*.
- As well as for the tag *audio* in the article *List of Internet stations*.

## 5. CONCLUSIONS

The present work shows how a tagging system can be integrated within a taxonomic system and, more precisely, how would a tag-based navigation affects on Wikipedia itself. On the one hand, a qualitative analysis over the presented prototype shows some new ways to navigate through articles:
- Pivot-browsing
- Popularity
- Filtering

On the other hand, a quantitative analysis shows a high number of new terms to appear for each article using tag information. This means additional terms for the search engine, as well as additional terms to navigate through.

Analyzing, in a general view, the navigation way by tags, it relies on a bottom-up classification rather than a top-down classification of the categories. This allows cross-navigating on the tags on a single level, while the categories require going up on the tree to look for new leaf categories.

Finally, the experiments carried out for this work show positive results, encouraging the use of a tagging system to improve navigation and search on Wikipedia. The main improvements of the use of tags over Wikipedia are:
- Providing new non-existing terms.
- Providing new ways to navigate through tags.
- Helping to improve the search engine.
- Discovering popular articles.